%% file: engrun2015_resonance_search.tex
\documentclass[twocolumn, showpacs, preprintnumbers,prd, superscriptaddress]{revtex4-1}
\pdfoutput=1

\usepackage{color}
\usepackage{cancel}
\usepackage{epstopdf}
\usepackage{epsfig}
\usepackage{booktabs}
 \usepackage{array,multirow}

\usepackage{amsmath, amssymb, mathrsfs}

\usepackage{graphicx}

\usepackage{dcolumn}

\usepackage{bm}

\usepackage[utf8]{inputenc}

\usepackage{hyperref}

\usepackage{xcolor}

\newcommand{\mevcc}{MeV/c$^2$}
\newcommand{\gevcc}{GeV/c$^2$}
\newcommand{\pos}{e^{+}}
\newcommand{\ele}{e^{-}}
\newcommand{\epem}{\pos\ele}
\newcommand{\emem}{\ele\ele}

\newcommand{\aprime}{A^\prime}

\begin{document}


	\title{Search for a Dark Photon in Electro-Produced $\epem$ Pairs 
    	   with the Heavy Photon Search Experiment at JLab}
    
    \input{authors}

    \date{\today}
    
    \begin{abstract}
        
        The Heavy Photon Search experiment took its first data in a 2015
        engineering run at the Thomas Jefferson National Accelerator Facility,
        searching for a prompt, electro-produced dark photon with a mass 
        between 19 and 81~\mevcc. A search for a resonance in the $\epem$
        invariant mass distribution, using 1.7 days (1170 nb$^{-1}$) of data, 
        showed no evidence of dark photon decays above the large QED background,
        confirming earlier searches and demonstrating the full functionality of
        the experiment. Upper limits on the square of the coupling of the dark
        photon to the Standard Model photon are set at the level of 
        6$\times$10$^{-6}$. Future runs with higher luminosity will explore new 
        territory. 

	\end{abstract}
	
    
    \maketitle
    
    \section{Introduction}
        
        The search for low-mass hidden sectors weakly coupled to the Standard 
        Model (SM) has received increased attention over the last
        decade~\cite{Hewett:2012ns,Essig:2013lka,Alexzzander:2016aln,
        Battaglieri:2017aum,Jaeckel:2010ni}.   Hidden sectors are motivated by
        the existence of dark matter, appear in myriad extensions of the SM, 
        and have been invoked to explain a wide variety of experimental anomalies.   

        A prototypical hidden sector consists of a spontaneously broken 
        ``hidden'' $U(1)'$ gauge symmetry, whose mediator is the 
        ``heavy photon'' or ``dark photon'', $\aprime$.  The heavy photon
        interacts with SM particles through kinetic mixing with the $U(1)_Y$ 
        (hypercharge) gauge boson~\cite{Holdom:1985ag,Galison:1983pa}, 
        resulting in the effective lagrangian density
        \begin{equation} \label{kmix} 
            {\cal L} \supset -\frac{\epsilon}{2\cos\theta_W} F'_{\mu\nu} F^{\mu\nu}_Y\,. 
        \end{equation}
        Here $\epsilon$ is a dimensionless coupling parameter, $\theta_W$ is the 
        Weinberg mixing angle, 
        $F'_{\mu\nu}=\partial_{\mu}\aprime_{\nu}-\partial_{\nu}\aprime_{\mu}$ 
        is the $U(1)'$ field strength, and similarly $F^{\mu\nu}_Y$ denotes the
        SM hypercharge $U(1)_Y$ field strength. This mixing generates an 
        interaction between the $\aprime$ and the SM photon at low energies,  
        allowing dark photons to be produced in charged particle interactions
        and, if sufficiently massive, to decay into pairs of charged particles 
        like $\epem$ or hidden-sector states.  The value of $\epsilon$ is 
        undetermined, but a value of $\epsilon^2 \sim 10^{-8}-10^{-4}$ is 
        natural if generated by quantum effects of heavier particles charged
        under $U(1)'$ and $U(1)_Y$.  If the SM forces unify in a Grand Unified 
        Theory, then $\epsilon^2 \sim 10^{-12}-10^{-6}$ is 
        natural~\cite{ArkaniHamed:2008qp,Baumgart:2009tn,Essig:2009nc}.  
        The mass of the $A'$, $m_{A'}$, is also undetermined, but the MeV-to-GeV
        mass scale has received much attention over the last decade as a 
        possible explanation for various anomalies related to dark matter
        interacting through the $A'$~\cite{ArkaniHamed:2008qn,Pospelov:2008jd,
        Finkbeiner:2007kk,Fayet:2004bw,Kaplinghat:2015aga} and for the 
        discrepancy between the observed and SM value of the muon anomalous 
        magnetic moment~\cite{Pospelov:2008zw,Bennett:2006fi,Davier:2010nc}.  
        Moreover, this mass range appears naturally in a few specific
        models~\cite{ArkaniHamed:2008qp,Cheung:2009fk,Baumgart:2009tn,
        Morrissey:2009ur,Essig:2009nc}. 

        The Heavy Photon Search (HPS) is an experiment utilizing the CEBAF
        accelerator at the Thomas Jefferson National Accelerator Facility (JLab)
        in Newport News, Virginia, USA. The experiment can explore a wide range
        of masses ($m_{\aprime}\sim 20-500$~\mevcc) and couplings 
        ($\epsilon^2 \sim 10^{-6} - 10^{-10}$), using both resonance search and
        separated vertex strategies. In this paper, results of a 
        resonance search from a Spring 2015 engineering run  
        using a 50~nA, 1.056~GeV electron beam impinging on a
        thin (0.125\%$X_{0}$) tungsten target are reported.  Electron interactions 
        with the 
        target nuclei could produce an $A'$ particle, which could subsequently 
        decay to an $\epem$ 
        pair~\cite{Bjorken:2009mm,Reece:2009un,Freytsis:2009bh}. A spectrometer,
        triggered by an electromagnetic calorimeter, measures the momenta and
        trajectories of this pair, allowing for the reconstruction of its
        invariant mass and decay position. The $\aprime$ would appear as a 
        narrow resonance, with a width set by the mass resolution, on top of a 
        smooth and wide distribution of background events from ordinary
        quantum electrodynamic (QED) processes.

        The cross section for  $\aprime$ production and subsequent decay to $\epem$      
        (``radiative $\aprime$ production'') scales with $\epsilon^2$ and is 
        directly proportional to the cross section for $\epem$ pair production
        from virtual photon bremsstrahlung 
        (``radiative trident production'')~\cite{Bjorken:2009mm}, so their
        yields are proportional. We assume the
        $\aprime$ only decays to $\epem$, as expected below the di-muon threshold
        if there are no invisible $\aprime$ decays.
        The measured $\epem$ yield, $dN/dm_{\aprime}$, is accounted for by
        the sum of trident and wide-angle bremsstrahlung (WAB) processes. Both
        radiative and Bethe Heitler diagrams contribute to trident production.
        WABs contribute if the photon converts and the resulting positron is detected
        along with the electron which has radiated.
        After accounting for the converted WABs, the trident yield is known.
        The fraction
        of all tridents which are radiative can be calculated, so the radiative trident
        yield is also determined, fixing the sensitivity of the search. 
        The experimental mass resolution impacts the experimental reach and is a
        critical input to the fits of the mass spectrum; it is calibrated by 
        measuring the invariant mass of M\o ller pairs, which have a unique 
        invariant mass for any given incident electron energy.

        The outline of the rest of the paper is as follows. 
        In Sec.~\ref{sec:detector}, we describe the experimental setup and the
        detector. Sec.~\ref{sec:selection} discusses the selection of the 
        events to maximize the $A'$ signal over the QED background.  
        Sec.~\ref{sec:resonance} describes the analysis of the resonance search,
        while Sec.~\ref{sec:results} presents the results.  Our conclusions are 
        presented in Sec.~\ref{sec:conclusions}. 
   
    \section{Detector Overview}\label{sec:detector}	
        
        The kinematics of $\aprime$ electro-production result in very 
        forward-produced heavy photons, which carry most of the beam energy and decay to
        highly-boosted $\epem$ pairs. To accept these decays, the HPS detector is 
        designed as a compact forward magnetic spectrometer, consisting of a silicon
        vertex tracker (SVT) placed in a vertical dipole magnetic field for momentum
        measurement and vertexing, and a PbWO$_4$ crystal electromagnetic calorimeter
       (ECal) for event timing and triggering. The SVT consists of six layers of
        detectors located in vacuum between 10 and 90 cm from the target, and
        arranged just above and below the ``dead zone'', a horizontal fan of intense
        flux from beam particles which have scattered or radiated in the target.
        Each
        layer consists of two silicon microstrip sensors with a small (50 or 100
        mrad) stereo angle for three dimensional position 
        determination~\cite{Adrian:2018}. The ECal  has 442 crystals and is situated downstream of the 
        tracker~\cite{Balossino:2016nly}. The ECal is split above and below the
        vacuum chamber which transports the beam towards the dump.
        
        HPS searches for a small signal above the much larger QED trident background, 
        so it must accumulate high statistics. This was
        accomplished using CEBAF's nearly continuous beam, 
        SVT and ECal readout with precision timing, and a high rate data acquisition
        system. The CEBAF accelerator provided a very stable beam with negligible
        halo, focused to a $\sim$100 $\mu$m spot at the 
        target~\cite{Baltzell:2016eee}. The SVT was read out using the APV25 ASIC operating at 41.333 MHz~\cite{French:2001xb} and triggered data from each sensor was sent to the SLAC ATCA-RCE readout system~\cite{Herbst:2016prn}. The ECal was read out with a 250 MHz 
        JLab FADC~\cite{Dong}. A custom trigger used the ECal information to
        select events consistent with coming from a high-energy $\epem$ pair.
        The data acquisition system could record events at rates up to 25 kHz with less than 15\% deadtime.
        
        The  analyzing magnet provided a field of 0.25 Tesla. The resulting
        SVT momentum resolution is $\delta p/p = 7$\% for beam energy electrons
        and is approximately constant for all momenta of 
        interest~\cite{Adrian:2018}. The ECal has an energy resolution 
        $\delta E/E = 5.7$\% at 0.5 GeV with significant energy and position
        dependence~\cite{Balossino:2016nly}. Using 
        information from the ECal and the SVT, we select $\epem$ pairs and
        reconstruct their invariant mass and vertex positions. This gives the
        experiment access to two regions of parameter space, comparatively
        large couplings using a traditional resonance search strategy, and very small 
        couplings using the distance from the target to the decay vertex to
        eliminate almost all of the prompt 
        trident background.      
        
        The HPS detector was installed and commissioned within the Hall B
        alcove at JLab early in the spring of 2015 and subsequently took its
        first data.  In total, 1170 nb$^{-1}$ of data was collected 
        (corresponding to 7.25 mC of integrated charge), equivalent to 
        1.7 days of continuous running. 
    
    \section{Event Selection}\label{sec:selection}
    
    	Searching for a heavy photon resonance requires accurate
        reconstruction of the $\epem$ invariant mass spectrum; 
        rejection of background events due to converted WAB events, non-radiative
        tridents from the Bethe-Heitler process, 
        and occasional accidental $\epem$ pairs; and efficient selection of $\aprime$ candidates.     
        Selecting $\aprime$ candidates is equivalent to selecting
        radiative tridents since they have identical kinematics for a
        given mass. In order to perform a blind search, the event selection  was 
        optimized using $\sim$10\% of the 2015 engineering run dataset.
        
        Heavy photon candidates are created from pairs of electron and positron
        tracks, one in each half of the SVT, each of which point to an energy  
        cluster in the ECal.  Each track must pass loose quality requirements
        and have a reconstructed momentum less than 75\% of the beam energy
        (0.788 \gevcc) to reject scattered beam electrons.  The background from 
        accidental pairs was reduced to less than 1\% by requiring the time between 
        the ECal clusters be less than 2 ns and the time between a track and the
        corresponding cluster be less than 5.8 ns.
        
        Heavy photons decay to highly boosted $\epem$ pairs, while the 
        recoiling electron is soft, scatters to large angles, and is usually
        undetected. Radiative tridents, having identical kinematics, comprise
        an irreducible background. The Bethe-Heitler diagram also contributes
        to trident production, and in fact dominates over the radiative process
        at all pair momenta. This background is minimized by requiring the
        momentum sum of the $\epem$ pair to be greater than 80\% of the beam
        energy (0.84 \gevcc), where the radiative tridents are peaked.
        
        The other significant source of background arises from converted WAB 
        events in which the bremsstrahlung photon is emitted at a large angle
        ($>15$ mrad), converts in the target, first or second layer of the SVT, and 
        gives rise to a detected positron in the opposite half of the detector 
        from the recoiling incoming electron. Although the fraction of such 
         WAB events that convert with this topology  is extremely low, 
        it is offset by the fact that 
        the bremsstrahlung rate is huge compared to the trident rate. This 
        results in converted WAB events making up roughly 30\% of our sample.  
        
        The converted WAB background was substantially reduced by applying additional 
        selection criteria. Since the conversion usually happens in the
        first layers of the silicon detector, requiring both tracks to have 
        hits in the first two layers of the SVT removes most of the converted WABs. 
        Requiring the transverse momentum asymmetry between the
        electron and positron be 
        $\frac{p_t(\ele)-p_t(\pos)}{p_t(\ele)+p_t(\pos)}<0.47$ and the 
        transverse distance of closest approach to the beam spot of the 
        positron track to be less 
        than 1.1 mm removes many of the remaining conversions. With all these cuts,
        contamination from converted WABs is reduced to 12\%.  
	
	    The composition of our event sample was checked by comparing the rates
        and distributions of several key variables (e.g total pair energy, 
        electron energy, positron energy, and invariant mass) between data and
        Monte Carlo (which included tridents, converted WABs, and accidental 
        background). We find that the data and MC are in reasonable agreement. 

    \section{Resonance Search}\label{sec:resonance}
       
\begin{figure}[th]
            \centering
            \includegraphics[width=.9\linewidth]{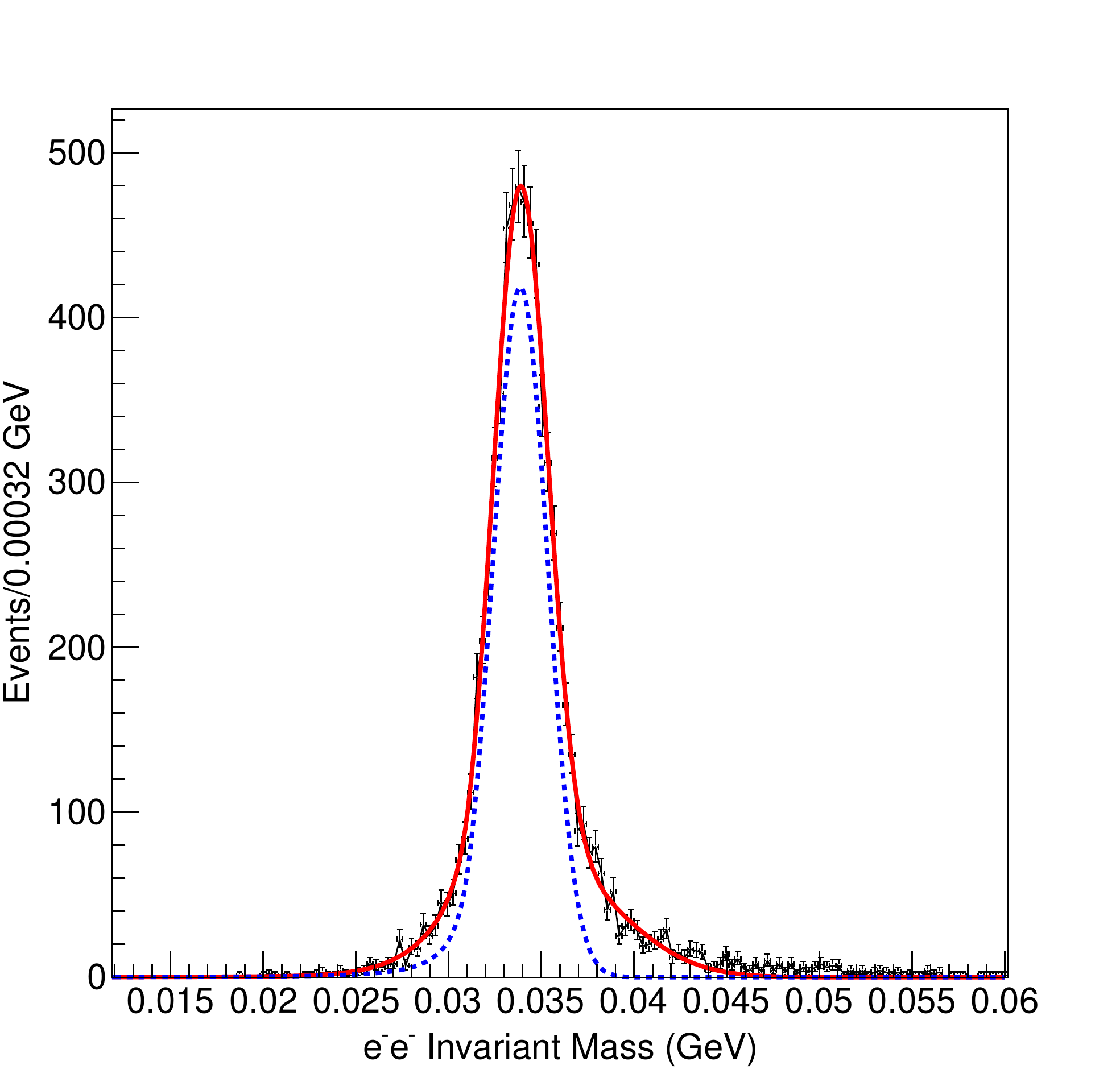}
            \caption{
                The M\o ller mass peak used to measure the mass resolution.
                The peak was fit with a Crystal Ball function plus a Gaussian
                for the tail at high mass. The $\sigma$ of the Crystal Ball 
                function was taken as the mass resolution. The overall fit is 
                in red; the core Crystal Ball in dashed blue.}
            \label{fig:mass_resolution}
        \end{figure}

        A heavy photon is expected to appear as a Gaussian-shaped resonance above the
        $\epem$ invariant mass spectrum, centered on the $\aprime$ mass and
        with a width, $\sigma_{m_{\aprime}}$, which characterizes the 
        experimental mass resolution.  M\o ller scattering events 
        ($\emem \rightarrow \emem$) are used to calibrate the $\aprime$ mass 
        scale and resolution.  Figure~\ref{fig:mass_resolution} shows the 
        measured M\o ller invariant mass, after a series of quality and 
        selection cuts.  For incident electrons of energy 1.056 GeV, we observe
        a M\o ller mass peak of 33.915 $\pm$ 0.043 MeV, within 1\% agreement of
        the expected mass of 34.1 MeV.  The M\o ller mass resolution predicted
        by Monte Carlo is 1.30 $\pm$ 0.02 MeV, in contrast with the observed 
        value of 1.61 $\pm$ 0.04 MeV.  We ascribe the difference to the fact 
        that our measured momentum resolution for beam energy electrons (7.03\%) 
        is significantly worse than predicted by Monte Carlo (5.9\%). Since
        the mass resolution scales directly with the momentum resolution, it 
        is underestimated in Monte Carlo by 19\%.  Consequently, we scale up
        the simulated $\aprime$ mass resolution by a factor of 1.19.
        The resulting parameterization of the mass resolution is an input to the 
        resonance search.
        
        Since the mass of a putative $\aprime$ is unknown a priori, the entire 
        $\epem$ invariant mass spectrum is scanned for any significant peaks.
        This search is performed in a broad mass window around each candidate 
        mass, repeated in 0.5 MeV steps between 19 and 81 MeV.  Searches above
        81 MeV are limited by both statistics and the incident electron beam
        energy. Within the
        window, which is 14$\sigma_{\aprime}$ wide below 39 MeV and 13$\sigma_{\aprime}$ wide
        between 39 and 81 MeV, the invariant mass distribution of $\epem$ 
        events is modeled using the probability distribution function
        \begin{equation}
            \resizebox{.9\linewidth}{!}{
                $P(m_{\epem}) = 
                        \mu \cdot \phi(m_{\epem} | m_{\aprime}, \sigma_{m_{\aprime}}) 
                        + B\cdot \exp(p(m_{\epem} | \mathbf{t}))$
            }
        \end{equation}
        where $m_{\epem}$ is the $\epem$ invariant mass, $\mu$ is the signal 
        yield, $B$ is the number of background events within the window, 
        $\phi(m_{\epem} | m_{\aprime}, \sigma_{m_{\aprime}})$ is a Gaussian probability 
        distribution describing the signal and $p(m_{\epem} | \mathbf{t})$ is a 
        Chebyshev polynomial of the first kind with coefficients 
        $\mathbf{t} = (t_{1}, ... t_{j})$ that is used to describe the 
        background shape. From optimization studies,
        a 5th (3rd) order Chebyshev polynomial was found to best 
        describe the background below (above) 39 MeV. Note that $m_{\aprime}$ 
        and $\sigma_{m_{\aprime}}$ are set 
        to the $\aprime$ mass hypothesis and expected experimental mass 
        resolution, respectively. Estimating the signal yield, the background
        normalization, and the background shape parameters within a window 
        is done with a binned maximum likelihood fit using a bin width of 0.05 MeV,
        which was found to have the lowest signal bias. A detailed discussion 
        of the procedures followed can be found in~\cite{Cowan:2010js}. Briefly,
        the log of the ratio of likelihoods for the background-only fit to that 
        of the best signal-plus-background fit provides a test statistic from 
        which the $p$-value can be calculated, giving the probability that the 
        observed signal is a statistical fluctuation.  The $p$-value is corrected 
        for the ``Look Elsewhere Effect'' (LEE) by performing simulated resonance 
        searches on 4,000 pseudo data sets. This relates the minimum $p$-value 
        seen in a given mass bin to the global probability of observing that 
        $p$-value in the search of the entire mass spectrum~\cite{Gross:2010qma}. 
    
    \section{Results}\label{sec:results}
    
\begin{figure}[h]
            \centering
            \includegraphics[width=.9\linewidth]{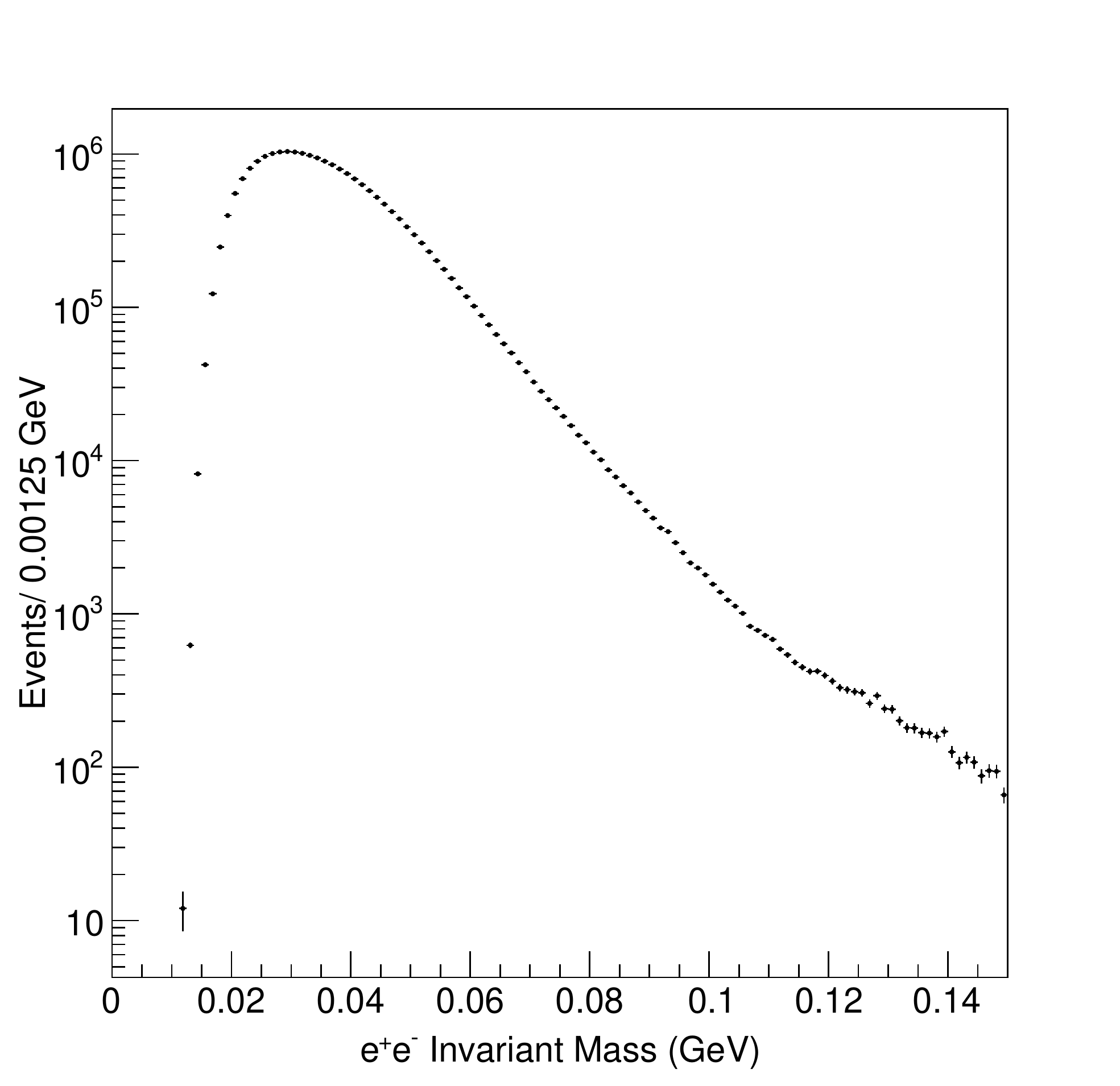}
            \caption{
                Distribution of $\epem$ invariant masses, events per 1.25 MeV 
                mass bin vs. mass.
            }
            \label{fig:mass}
        \end{figure}

        A search for a resonance in the $\epem$ invariant mass spectrum, 
        shown in Figure~\ref{fig:mass}, between 19 MeV and 81 MeV found no 
        evidence of an $\aprime$ signal. The most significant signal was 
        observed at 37.7 MeV and has a local $p$-value of 0.17\%.  After 
        accounting for the LEE correction, the most significant $p$-value is 
        found to have a global $p$-value of 17\% corresponding to less than
        2$\sigma$ in significance. Since no significant signals were found, 
        a  95\% C.L. upper limit is set, power-constrained~\cite{Cowan:2011an}
        to the expected limit. 

        \begin{figure}[h]
            \centering
            \includegraphics[width=\linewidth]{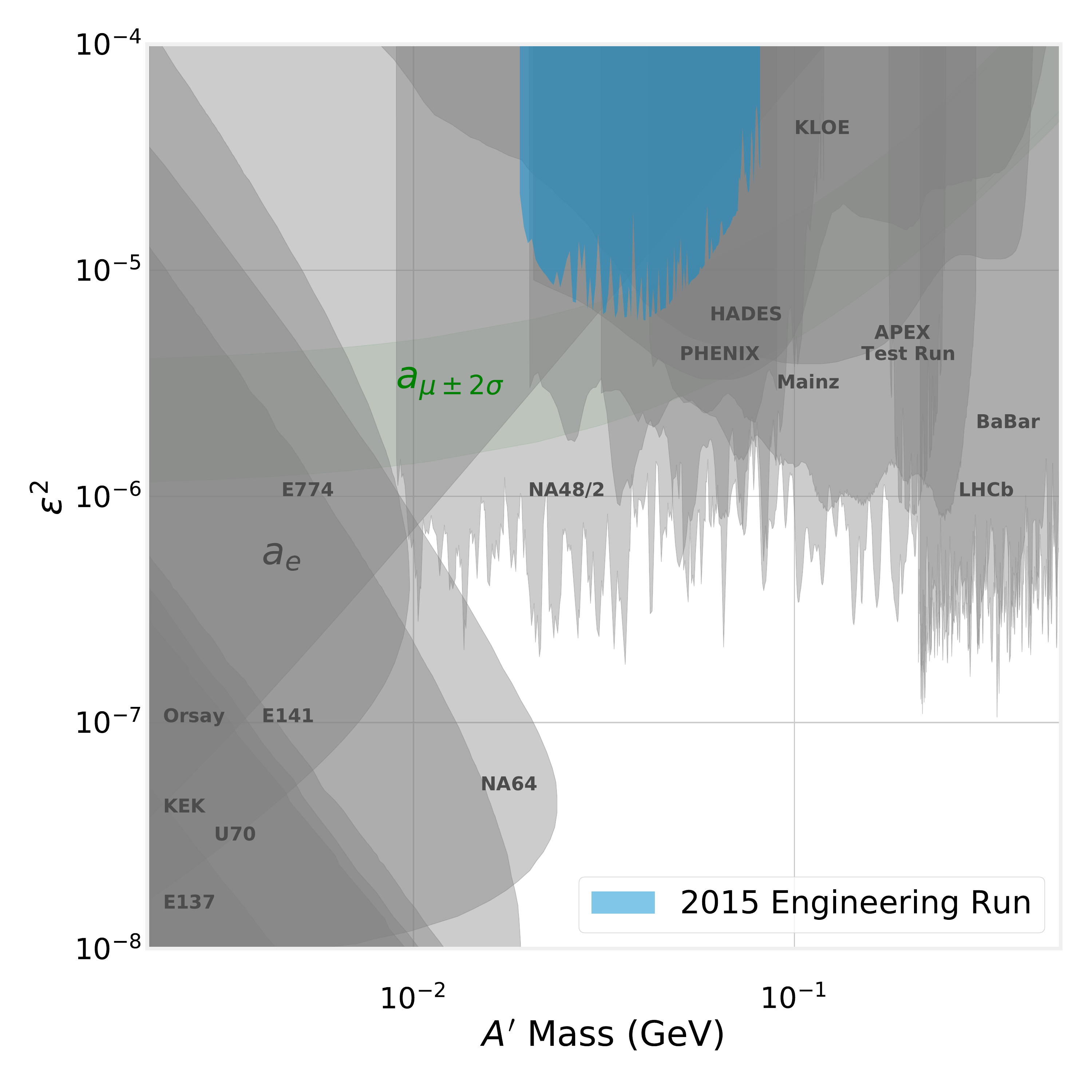}
            \caption{
             The 95\% C.L. power-constrained~\cite{Cowan:2011an} upper limits on 
             $\epsilon^2$ versus $\aprime$ mass obtained in this analysis. A 
             limit at the level of 6$\times$ 10$^{-6}$ is set. Existing 
             limits from beam dump~\cite{Bjorken:1988as, riordan1987, bross1991, konaka1986,
             davier1989, Bjorken:2009mm, andreas2012, Blumlein:1990ay, Blumlein:1991xh}, 
             collider~\cite{Reece:2009un, Aubert:2009cp, Babusci:2012cr, Archilli:2011zc, Aaij:2017rft} 
             and fixed target experiments~\cite{Abrahamyan:2011gv, Merkel:2014avp,
            Agakishiev:2013fwl, Batley:2015lha} are also shown. 
             The region labeled ``$a_e$'' is an exclusion based 
             on the electron $g-2$~\cite{PhysRevLett.106.080801, Aoyama:2012wj, PhysRevLett.100.120801, Davoudiasl:2012ig}
             . The green band labeled ``$a_{\mu} \pm 2\sigma$''
             represents the region that an $A'$ can be used to explain the discrepancy 
             between the measured and calculated muon anomalous magnetic moment~\cite{Pospelov:2008zw, Bennett:2006fi}.
            }          
            \label{fig:epsilon_upper_limit}
        \end{figure}

        The proportionality between $\aprime$ and radiative trident 
        production allows the normalization of the $\aprime$ rate to the 
        measured rate of trident production~\cite{Bjorken:2009mm}.  This leads
        to a relation that allows the signal upper limit, $S_{\text{up}}$, to be 
        related to the $\aprime$ coupling strength as 
        \begin{equation} \label{eps_coupling}
            \epsilon^2 = \left (\frac{S_{\text{up}}/m_{A'}}{
                    f\Delta B/\Delta m} \right) 
                    \left(\frac{2 N_{eff} \alpha}{3 \pi} \right)
        \end{equation}
        where $N_{eff}$ is the number of decay channels kinematically accessible
        (=1 for HPS searches below the dimuon threshold), $\Delta B/\Delta m$
        is the number of background events per MeV, $\alpha$ is the fine structure
        constant and $f$ = 8.5\% is the 
        fraction of radiative trident events comprising the background.  Using
        equation \ref{eps_coupling}, the limits on $\epsilon$ set by HPS are
        shown on Figure \ref{fig:epsilon_upper_limit}. 
        
        The reach shown in Figure \ref{fig:epsilon_upper_limit} includes all 
        statistical and  systematic uncertainties. The main systematic 
        uncertainties on the signal yields arise from the uncertainty in the mass
        resolution (3\%) and biases observed in the fit due to the background 
        and signal parameterization (1.3-1.5\%, depending on mass). When scaling
        the extracted signal yield upper limits to a limit on $\epsilon$, the
        primary systematic uncertainty in the radiative fraction is due to 
        the unknown composition of the final $\epem$ sample (7\%).  Many
        other possible sources of systematic uncertainty were investigated and
        accounted for but contribute negligibly to the result.

	\section{Conclusion} \label{sec:conclusions}
    
    	A resonance search for a heavy photon with a mass between 19 and 81 MeV
        which decays to an $\epem$ pair was performed.  A search for a resonance in
        the $\epem$ invariant mass spectrum yielded no significant excess and 
        established upper limits on the square of the coupling at the level of
        $6\times10^{-6}$, confirming results of earlier searches. While not
        covering new territory in this short engineering run, this search did
        establish that HPS operates as designed and will, with future running,
        extend coverage for $\epsilon^2$ below the level of 10$^{-6}$. Coverage
        of unexplored parameter space at smaller values of the coupling will be
        possible from a search for events with displaced vertices.

    \section{Acknowledgments}
 
        The authors are grateful for the outstanding efforts of the Jefferson 
        Laboratory Accelerator Division and the Hall B engineering group in 
        support of HPS. The research reported here is supported by the U.S.
        Department of Energy Office of Science, Office of Nuclear Physics, 
        Office of High Energy Physics, the French Centre National de la 
        Recherche Scientifique, 
        United Kingdom's Science and Technology Facilities Council (STFC),
        the Sesame project HPS@JLab funded by the French region Ile-de-France 
        and the Italian Istituto Nazionale di Fisica Nucleare. Jefferson Science
        Associates, LLC, operates the Thomas Jefferson National Accelerator
        Facility for the United States Department of Energy under Contract
        No. DE-AC05-060R23177. 
    
    \bibliography{bibliography}

\end{document}

%% file: authors.tex
%
%
%

\newcommand*{\SACLAY}{IRFU, CEA, Universit\'e Paris-Saclay, F-91191 Gif-sur-Yvette, France}
\newcommand*{\SACLAYindex}{1}
\newcommand*{\INFNGE}{INFN, Sezione di Genova, 16146 Genova, Italy}
\newcommand*{\INFNGEindex}{2}
\newcommand*{\INFNTUR}{INFN, Sezione di Torino, 10125 Torino, Italy}
\newcommand*{\INFNTURindex}{21}
\newcommand*{\ORSAY}{Institut de Physique Nucl\'eaire, CNRS-IN2P3,
                    Univ. Paris-Sud, Universit\'e Paris-Saclay, 91406 Orsay, France}
\newcommand*{\ORSAYindex}{22}
\newcommand*{\UNH}{University of New Hampshire, Durham, New Hampshire 03824, USA}
\newcommand*{\UNHindex}{27}
\newcommand*{\NSU}{Norfolk State University, Norfolk, Virginia 23504, USA}
\newcommand*{\NSUindex}{28}
\newcommand*{\ODU}{Old Dominion University, Norfolk, Virginia 23529, USA}
\newcommand*{\ODUindex}{30}
\newcommand*{\ROMAII}{Universit\`a di Roma Tor Vergata, 00133 Rome Italy}
\newcommand*{\ROMAIIindex}{32}
\newcommand*{\JLAB}{Thomas Jefferson National Accelerator Facility, Newport News, Virginia 23606, USA}
\newcommand*{\JLABindex}{36}
\newcommand*{\GLASGOW}{University of Glasgow, Glasgow G12 8QQ, United Kingdom}
\newcommand*{\GLASGOWindex}{39}
\newcommand*{\WM}{College of William \& Mary, Williamsburg, Virginia 23187, USA}
\newcommand*{\WMindex}{42}
\newcommand*{\YEREVAN}{Yerevan Physics Institute, 375036 Yerevan, Armenia}
\newcommand*{\YEREVANindex}{43}
\newcommand*{\PERIMETER}{PerimeteVr Institute, Ontario, Canada N2L 2Y5}
\newcommand*{\PERIMETERindex}{44}
\newcommand*{\SASSARI}{Universit\`a di Sassari, 07100 Sassari, Italy}
\newcommand*{\SASSARIindex}{45}
\newcommand*{\FNAL}{Fermi National Accelerator Laboratory, Batavia, IL 60510, USA}
\newcommand*{\FNALindex}{46} 
\newcommand*{\CATANIA}{INFN, Sezione di Catania, 95123 Catania, Italy}
\newcommand*{\CATANIAindex}{47}
\newcommand*{\STONYBROOK}{C.~N.~Yang Institute for Theoretical Physics, 
                          Stony Brook University, Stony Brook, NY 11794, USA}
\newcommand*{\STONYBROOKindex}{48}
\newcommand*{\SLAC}{SLAC National Accelerator Laboratory, Stanford University, Stanford, CA 94309, USA}
\newcommand*{\SLACindex}{49}
\newcommand*{\UCSC}{Santa Cruz Institute for Particle Physics, University of California, Santa Cruz, CA 95064, USA}
\newcommand*{\UCSCindex}{50}
\newcommand*{\PADOVA}{Universit\`a di Padova, 35122 Padova, Italy} 
\newcommand*{\PADOVAindex}{51}
\newcommand*{\INFNPA}{INFN, Sezione di Padova, 16146 Padova, Italy}
\newcommand*{\INFNPAindex}{52}
\newcommand*{\IDAHO}{Idaho State University, Pocatello, ID, 83209, USA}
\newcommand*{\IDAHOindex}{53}
\newcommand*{\INFNROMA}{INFN, Sezione di Roma Tor Vergata, 00133 Rome, Italy}
\newcommand*{\INFNROMAindex}{54}
\newcommand*{\INFNSUD}{INFN, Laboratori Nazionali del Sud, 95123 Catania, Italy}
\newcommand*{\INFNSUDindex}{55}

\author{P.~H. Adrian}\affiliation\SLAC
\author{N.~A.~Baltzell}\affiliation\JLAB
\author{M.~Battaglieri}\affiliation\INFNGE
\author{M.~Bond\'i}\affiliation\CATANIA
\author{S.~Boyarinov}\affiliation\JLAB
\author{S.~Bueltmann}\affiliation\ODU
\author{V.~D.~Burkert}\affiliation\JLAB
\author{D.~Calvo}\affiliation\INFNTUR
\author{M.~Carpinelli}\affiliation\SASSARI\affiliation\INFNSUD
\author{A.~Celentano}\affiliation\INFNGE
\author{G.~Charles}\affiliation\ORSAY
\author{L.~Colaneri}\affiliation\ROMAII\affiliation\INFNROMA
\author{W.~Cooper}\affiliation\FNAL
\author{C.~Cuevas}\affiliation\JLAB
\author{A.~D'Angelo}\affiliation\ROMAII\affiliation\INFNROMA
\author{N.~Dashyan}\affiliation\YEREVAN
\author{M.~De~Napoli}\affiliation\CATANIA
\author{R.~De~Vita}\affiliation\INFNGE
\author{A.~Deur}\affiliation\JLAB
\author{R.~Dupre}\affiliation\ORSAY
\author{H.~Egiyan}\affiliation\JLAB
\author{L.~Elouadrhiri}\affiliation\JLAB
\author{R.~Essig}\affiliation\STONYBROOK
\author{V.~Fadeyev}\affiliation\UCSC
\author{C.~Field}\affiliation\SLAC
\author{A.~Filippi}\affiliation\INFNTUR
\author{A.~Freyberger}\affiliation\JLAB
\author{M.~Gar\c{c}on}\affiliation\SACLAY
\author{N.~Gevorgyan}\affiliation\YEREVAN
\author{F.~X.~Girod}\affiliation\JLAB
\author{N.~Graf}\affiliation\SLAC
\author{M.~Graham}\affiliation\SLAC
\author{K.~A.~Griffioen}\affiliation\WM
\author{A.~Grillo}\affiliation\UCSC
\author{M.~Guidal}\affiliation\ORSAY
\author{R.~Herbst}\affiliation\SLAC
\author{M.~Holtrop}\affiliation\UNH
\author{J.~Jaros}\affiliation\SLAC
\author{G.~Kalicy}\affiliation\ODU
\author{M.~Khandaker}\affiliation\IDAHO
\author{V.~Kubarovsky}\affiliation\JLAB
\author{E.~Leonora}\affiliation\CATANIA
\author{K.~Livingston}\affiliation\GLASGOW
\author{T.~Maruyama}\affiliation\SLAC
\author{K.~McCarty}\affiliation\UNH
\author{J.~McCormick}\affiliation\SLAC
\author{B.~McKinnon}\affiliation\GLASGOW
\author{K.~Moffeit}\affiliation\GLASGOW
\author{O.~Moreno}\email[Corresponding author, email:]{omoreno@slac.stanford.edu}\affiliation\SLAC\affiliation\UCSC
\author{C.~Munoz~Camacho}\affiliation\ORSAY
\author{T.~Nelson}\affiliation\SLAC
\author{S.~Niccolai}\affiliation\ORSAY
\author{A.~Odian}\affiliation\SLAC
\author{M.~Oriunno}\affiliation\SLAC
\author{M.~Osipenko}\affiliation\INFNGE
\author{R.~Paremuzyan}\affiliation\UNH
\author{S.~Paul}\affiliation\WM
\author{N.~Randazzo}\affiliation\CATANIA
\author{B.~Raydo}\affiliation\JLAB
\author{B.~Reese}\affiliation\SLAC
\author{A.~Rizzo}\affiliation\ROMAII\affiliation\INFNROMA
\author{P.~Schuster}\affiliation\SLAC\affiliation\PERIMETER
\author{Y.~G.~Sharabian}\affiliation\JLAB
\author{G.~Simi}\affiliation\PADOVA\affiliation\INFNPA
\author{A.~Simonyan}\affiliation\ORSAY
\author{V.~Sipala}\affiliation\SASSARI\affiliation\INFNSUD
\author{D.~Sokhan}\affiliation\GLASGOW
\author{M.~Solt}\affiliation\SLAC
\author{S.~Stepanyan}\affiliation\JLAB
\author{H.~Szumila-Vance}\affiliation\JLAB\affiliation\ODU
\author{N.~Toro}\affiliation\SLAC\affiliation\PERIMETER
\author{S.~Uemura}\affiliation\SLAC
\author{M.~Ungaro}\affiliation\JLAB
\author{H.~Voskanyan}\affiliation\YEREVAN
\author{L.~B.~Weinstein}\affiliation\ODU
\author{B.~Wojtsekhowski}\affiliation\JLAB
\author{B.~Yale}\affiliation\UNH